\documentstyle[12pt]{article}

\newcounter{eq}
\newcounter{sc}

\def\overleftrightarrow#1{\vbox{\ialign{##\crcr
 $\leftrightarrow$\crcr\noalign{\kern-1pt\nointerlineskip}
 $\hfil\displaystyle{#1}\hfil$\crcr}}}

\newlength{\minitwocolumn}
\begin{document}

\begin{flushright}
DPUR/TH/31\\
July, 2012\\
\end{flushright}
\vspace{20pt}

\pagestyle{empty}
\baselineskip15pt

\begin{center}
{\large\bf Remarks on Two Gamma Ray Lines from the Inner Galaxy
\vskip 1mm }

\vspace{20mm}
Ichiro Oda \footnote{E-mail address:\ ioda@phys.u-ryukyu.ac.jp
}

\vspace{5mm}
           Department of Physics, Faculty of Science, University of the 
           Ryukyus,\\
           Nishihara, Okinawa 903-0213, Japan.\\

\end{center}


\vspace{5mm}
\begin{abstract}
Monochromatic gamma-ray lines are thought to be the smoking gun signal of the annihilation
or decay of dark matter since they do not suffer from deflection or absorption on galactic
scales. A recent claim on strong evidence for two gamma-ray lines from the inner galaxy
suggests that two-body final states might be one photon plus a Z boson or one photon plus 
a Higgs boson. In this study, we investigate which final state is more possible 
by analyzing the energy resolution of the Fermi-LAT. It is concluded that the former case, i.e. 
one photon plus a Z boson is more plausible than the latter one, i.e. one photon and a Higgs boson
since in the latter case the mass of dark matter particle shows tension with a constraint coming from
the energy resolution of the Fermi-LAT.   
\end{abstract}

\newpage
\pagestyle{plain}
\pagenumbering{arabic}


\rm

There is overwhelming evidence that dark matter pervades our universe undoubtedly, but
the identity and nature of dark matter have remained elusive thus far \cite{Bertone}. 
The possible connection with proposed extensions beyond the Standard Model (SM) of elementary particles,
currently being seeked for at Large Hadron Collider (LHC), makes the identification of dark matter one of the
highest priority purposes in particle physics and cosmology. 
Among a lot of candidates of dark matter particles (denoted as $\chi$ in what follows), 
the fact that the annihilation cross-section at the electro-weak scale yields the appropriate relic 
abundance without any artificial fine-tuning has made weakly interacting massive particles (WIMPs) 
natural dark matter particles \cite{Jungman}.  

There are basically three kinds of methods for detecting dark matter particles, which are collider
searches, direct and indirect detections.  In particular, in the indirect method, one searches
for the dark matter annihilation or decay products in cosmic rays which include gamma-rays, electrons,
positrons and neutrinos etc. Most of interests in the indirect detection have been recently focused
on gamma-rays generated by dark matter annihilation from regions in the surrounding universe with a
high density of dark matter such as the galactic center, dwarf spheroidal galaxies and galaxy clusters. 
Here the lore is that only the dark matter annihilation can produce monochromatic photon lines whereas 
all standard cosmic phenomena we are aware of have continuous spectra which can be well approximated 
by a power law. To put differently, an observation of gamma-ray lines from the galactic center etc. 
can be regarded as a smoking gun signal of dark matter. 

However, the world is not so simple and there is a catch. Dark matter is by definition 'dark', so
it does not couple to photons directly at the tree-level. Thus, the dark matter annihilation into photons is 
induced by loop effects whose rate is in general orders of magnitudes smaller (typically, by the
factor $\alpha^2$ with $\alpha$ being the fine-structure constant) than the annihilation into other
SM particles. For this reason, most of models treating with dark matter do not account for 
gamma-ray lines at an observable level although there exist a handful of models where the dark
matter annihilation into photons is enhanced by some mechanism and consequently the resulting lines
could be observed. Another reason why it is difficult to observe the photon lines is that current
detectors have an energy resolution not better than $10\%$, thereby implying that the signal might
be smeared or drowned in the continuous background spectra.
  
In a recent paper by Weniger \cite{Weniger}, it is claimed that a monochromatic gamma-ray line is present at 
$E_\gamma \approx 130 GeV$ in the data collected during about last 4 years by the Fermi satellite \cite{Fermi}
with a local (global, i.e. after taking account of the look-elsewhere effect) significance of 
$4.6 \sigma \ (3.3 \sigma)$.\footnote{The paper by Weniger has subsequently given rise to much activity on this
subject \cite{Profumo}.} 
When interpreted as dark matter particles annihilating into a pair of photons \cite{Bergstrom}, 
the dark matter mass becomes $M_\chi \approx 130 GeV$ and a partial annihilation cross-section
takes the value $\langle \sigma v \rangle_{\chi \chi \rightarrow \gamma \gamma} \approx 10^{-27} cm^3/s$, which is a typical
cross-section value for weak interaction, and about $1/10$ of the total annihilation cross-section
expected if dark matter is a thermal relic from the big bang. However, we should mention that there is some 
tension between this finding of the gamma-ray line and the result of a line-research analysis by the Fermi-LAT 
collaboration \cite{Fermi} even if this tension might stem from a different method of data-analysis.

More recently, using the Fermi-LAT data again, Su and Finkbeiner examine the diffuse $80-200 GeV$ emission 
in the inner galaxy, and claim that a pair of gamma-ray lines, near $110 GeV$ and $130 GeV$, could provide us 
with a marginally better fit than a single line \cite{Su}. An appealing point of their study is that they analyzed 
photons within $5^\circ$ of galaxy center while the Fermi-LAT collaboration puts constraints on line emission coming from 
regions outside $5^\circ$ of galaxy center, so there is no conflict between results obtained by both the groups.   

Moreover, Su and Finkbeiner also mention that the pair of gamma-ray lines shows a WIMP of mass $127 GeV$ annihilating 
to $\gamma \gamma$ and $\gamma Z$ \cite{Bergstrom2}. Even more interestingly, they say that the pair of lines could be compatible 
with the "Higgs in Space" scenario \cite{Jackson} where a WIMP of mass $141 GeV$ annhilates into a photon $\gamma$ and a Higgs particle
$h$. It is worthwhile to emphasize that if the latter scenario were really true, the indirect detection method
of dark matter could have confirmed the existence of the Higgs particle of $125 GeV$ before LHC with the local significance $5 \sigma$! 

In this short article, assuming that a correct picture obtained from the Fermi-LAT data is not a single photon line like Weniger
but two lines like Su and Finkbeiner, we wish to ask ourselves the following interesting question: Is the "Higgs in Space" scenario able to be
confirmed through the analysis of the Fermi-LAT data? In order to answer it, the key idea lies in an observation 
that the energy resolution of the Fermi-LAT is $10 \%$, i.e. $\frac{\Delta E_\gamma}{E_\gamma} = 0.1$ and this
uncertainty leads to a condition on value of the dark matter mass. Then, it turns out that the value of the dark matter particle mass 
in the "Higgs in Space" scenario is in tension with this condition, so we cannot definitively state that the pair of gamma-ray lines is 
compatible with the "Higgs in Space" scenario. 

Let us start with the following physical situation:
Two dark matter particles which annihilate into $\gamma + Y$ (where $Y$ describes a photon $\gamma$, a Z boson $Z$ or 
a Higgs particle $h$) move with non-relativistic velocities $v \approx 10^{-3} c$. Then, it is easy to see that
the energy and momentum conservation laws provide the photon with energy
\begin{eqnarray}
E_\gamma = M_\chi ( 1 - \frac{M_Y^2}{4 M_\chi^2} ).
\label{Energy}
\end{eqnarray}
In particular, for the annihilation process $\chi + \chi \rightarrow \gamma + \gamma$, we have two photons with energy
equal to the dark matter mass $E_\gamma = M_\chi$, which shows a monochromatic $\gamma \gamma$ line with 
a small energy spread owing to the Doppler effect. 
A similar gamma-ray peak is expected to be observed for the $\gamma Z$ final state as well as the $\gamma h$ one,
but in these cases the intrinsic width of the $Z$ boson and the Higgs particle has to be considered.

One advantage of the satellite experiments like the Fermi-LAT compared to the terrestrial ones is the excellent energy
resolution possible, by which one can distinguish a line of $\gamma Z$ final state from that of $\gamma \gamma$ (or  $\gamma h$). 
If not a single but two lines were observed, a comparison between the two lines would give us useful information on a underlying 
theory behind dark matter physics. For instance, a comparison of line strenghts would provide information on the
cross-sections.

Furthermore, it is worthwhile to point out that the energy resolution $\varepsilon \equiv \frac{\Delta E_\gamma}{E_\gamma}$
gives us the condition that the two lines are separable. In the case of $Y = \{\gamma, Z\}$, the two lines take the photon energy
\begin{eqnarray}
E_{1\gamma} = M_\chi, \hspace{2.0em}  E_{2\gamma} = M_\chi ( 1 - \frac{M_Z^2}{4 M_\chi^2} ).
\label{Lines1}
\end{eqnarray}
The difference of energy between the two lines $\delta E_\gamma$ and its relative ratio $\delta$ are 
respectively defined as
\begin{eqnarray}
\delta E_\gamma &\equiv& E_{1\gamma} - E_{2\gamma} = \frac{M_Z^2}{4 M_\chi},  \nonumber\\
\delta &\equiv& \frac{\delta E_\gamma}{E_\gamma}.
\label{Difference1}
\end{eqnarray}
Now let us note that the condition for distinguishing the two photon lines is given by
\begin{eqnarray}
\varepsilon  \ll \delta \Longleftrightarrow   \varepsilon \ll \frac{M_Z^2}{4 M_\chi E_\gamma}.
\label{Condition1}
\end{eqnarray}
In other words, to discriminate the two photon lines with precision, the energy
resolution of a detector must be sufficiently less than the ratio of the energy difference of two peaks
to the photon energy observed in the detection. 

Since it is now natural to choose $E_\gamma = E_{1\gamma} = M_\chi$, the condition (\ref{Condition1})
leads to the constraint on the dark matter mass
\begin{eqnarray}
M_\chi \ll \frac{M_Z}{\sqrt{4 \varepsilon}}.
\label{DM mass1}
\end{eqnarray}
The Fermi-LAT has the energy resolution $\varepsilon = 0.1$ and the Z boson has mass $M_Z = 91 GeV$,
so this constraint reads\footnote{We nelect a contribution from the velocity of dark matter since
it only gives rise to a slight modification of the dark matter mass.}   
\begin{eqnarray}
M_\chi \ll 144 GeV.
\label{Limit1}
\end{eqnarray}
In the paper by Su et al. \cite{Su}, the dark matter mass is calculated to be $M_\chi = 127 GeV$ whose
value is consistent with this constraint (\ref{Limit1}).

Next, let us move on to a case of different gamma-ray lines $Y = \{Z, h \}$ where the two lines take 
the photon energy
\begin{eqnarray}
E_{1\gamma} = M_\chi ( 1 - \frac{M_Z^2}{4 M_\chi^2} ), \hspace{2.0em}  
E_{2\gamma} = M_\chi ( 1 - \frac{M_h^2}{4 M_\chi^2} ).
\label{Lines2}
\end{eqnarray}
Then, the difference between the two lines and its relative ratio are respectively given by
\begin{eqnarray}
\delta E_\gamma &\equiv& E_{1\gamma} - E_{2\gamma} = \frac{M_h^2 - M_Z^2}{4 M_\chi},  \nonumber\\
\delta &\equiv& \frac{\delta E_\gamma}{E_\gamma} = \frac{M_h^2 - M_Z^2}{4 M_\chi E_\gamma}.
\label{Difference2}
\end{eqnarray}
In this case as well, taking $E_\gamma = E_{1\gamma}$, the condition $\varepsilon  \ll \delta$
leads to the constraint on the dark matter mass
\begin{eqnarray}
M_\chi \ll \sqrt{\frac{M_h^2 + ( \varepsilon - 1 ) M_Z^2}{4 \varepsilon}}.
\label{DM mass2}
\end{eqnarray}
Using the fact that $\varepsilon = 0.1$ and the Higgs boson has mass $M_h = 125 GeV$,
one arrives at 
\begin{eqnarray}
M_\chi \ll 143 GeV.
\label{Limit2}
\end{eqnarray}
In this case, the dark matter mass is calculated to be $M_\chi = 141 GeV$ \cite{Su} which
is almost the same size as the upper bound $143 GeV$ in (\ref{Limit2}). 
In this sense, the calculated value $M_\chi = 141 GeV$ is in tension with the constraint (\ref{Limit2}), 
so an interpretation of the two photon lines as dark matter annihilating into $\gamma Z$ and $\gamma h$ 
is dubious. Incidentally, in the extreme situation $E_\gamma = M_\chi$, Eq. (\ref{Limit2})
reads $M_\chi \ll 135 GeV$ which is obviously inconsistent with $M_\chi = 141 GeV$. 
The only way to have a consistent result is to improve the energy resolution of the detector. 
For instance, if we can have the more excellent energy resolution $\varepsilon = 0.05 \ (\varepsilon = 0. 01)$, 
the constraint (\ref{DM mass2}) on the dark mass is reduced to the form $M_\chi \ll 197 GeV \ (M_\chi \ll 430 GeV)$ 
which is compatible with $M_\chi = 141 GeV$.

Now let us comment on species of the dark matter particles. It is natural to suppose that
the dark matter annihilation occurs in the s-wave state. Then, if final states of the 
two photon lines are found to be $Y = \{\gamma, Z\}$, the dark matter particle must be
a scalar boson, Majorana fermion or Dirac fermion from the relation between statistics 
and spin. On the other hand, if final states turn out to be $Y = \{Z, h \}$, 
the dark matter particle should be a Dirac fermion or a vector boson. In particular, note 
that the Landau-Yang theorem \cite{Landau} requires us that a vector particle cannot annihilate into 
two photons since the spatial part of wave function must be symmetric.\footnote{The multiplet 
with the total $1$ spin is always anti-symmetric when composed from two photons.} 
Furthermore, note that in this context the Dirac fermion has an interesting feature
in that if the dark matter particle were a Dirac fermion, three photon lines could be
observed as long as some (unknown) symmetry does not forbid the three annihilation processes
and the processes are allowed kinematically.   

In this article, it has been shown that an interpretation of two gamma-ray lines 
found in Ref. \cite{Su}, as the dark matter annihilation into $Z$ and $h$ in the
"Higgs in Space" scenario shows tension with the energy resolution of the Fermi-LAT.
On the other hand, another interpretation of them as the annihilation into 
$\gamma$ and $Z$ is consistent with it. Moreover, we have proposed that in order to
advocate the "Higgs in Space" scenario, the energy resolution needs to be improved 
in future. As a future work, we wish to take into consideration a issue of the velocity
of dark matter in the framework developed recently by the author \cite{Oda}.

\begin{flushleft}
{\bf Acknowledgements}
\end{flushleft}

This work is supported in part by the Grant-in-Aid for Scientific 
Research (C) No. 22540287 from the Japan Ministry of Education, Culture, 
Sports, Science and Technology.



\begin{thebibliography}{99}

\bibitem{Bertone}
G. Bertone (ed.), {"Particle Dark Matter: Observations, Models and Searches", 
Cambridge University Press, 2010.}

\bibitem{Jungman}
For review and references, see for instance, G. Jungman, M. Kamionkowski and K. Griest, 
{Phys. Rept. {\bf 267} (1996) 195}; G. Bertone, D. Hooper and J. Silk, {Phys. Rept. 
{\bf 405} (2005) 279.}

\bibitem{Weniger}
C. Weniger, {arXiv:1204.2797 [hep-ph].} 

\bibitem{Fermi}
M. Ackermann et al. [Fermi-LAT Collaboration], {arXiv:1205.2739 [astro-ph.HE].} 

\bibitem{Profumo}
S. Profumo and T. Linden, {arXiv:1204.6047 [astro-ph.HE]};
A. Ibarra, S. Lopez Gehler and M. Pato, {arXiv:1205.0007 [hep-ph]};
E. Tempel, A. Hektor and M. Raidal, {arXiv:1205.1045 [hep-ph]};
E. Dudas, Y. Mambrini, S. Pokorski and A. Romagnoni, {arXiv:1205.1520 [hep-ph]};
J. M. Cline, {arXiv:1205.2688 [hep-ph]};
A. Boyarsky, D. Malyshev and O. Ruchayskiy, {arXiv:1205.4700 [astro-ph.HE]}; 
L. Bergstrom,  {arXiv:1205.4882 [astro-ph.HE]};
A. Rajaraman, T. M. P. Tait and D. Whiteson, {arXiv:1205.4723 [hep-ph]};
K. -Y. Choi and O. Seto, {arXiv:1205.3276 [hep-ph]};
B. Kyae and J. -C. Park, {arXiv:1205.4151 [hep-ph]};
H. M. Lee, M. Park and W. -I. Park, {arXiv:1205.4675 [hep-ph]};
B. S. Acharya, G. Kane, P. Kumar, R. Lu and B. Zheng, {arXiv:1205.5789 [hep-ph]};
M. R. Buckley and D. Hooper, {arXiv:1205.6811 [hep-ph]};
X. Chu, T. Hambye, T. Scarna and M. H. G. Tytgat, {arXiv:1206.2279 [hep-ph]};
D. Das, U. Ellwanger and P. Mitropoulos, {arXiv:1206.2639 [hep-ph]};
Z. Kang, T. Li, J. Li and Y. Liu, {arXiv:1206.2863 [hep-ph]};
N. Weiner and I. Yavin, {arXiv:1206.2910 [hep-ph]};
L. Feng, Q. Yuan and Y. -Z. Fan, {arXiv:1206.4758 [astro-ph.HE]};
W. Buchmuller and M. Garny, {arXiv:1206.7056 [hep-ph]};
T. Cohen, M. Lisanti, T. R. and J. G. Wacker, {arXiv:1207.0800 [hep-ph].}

\bibitem{Bergstrom}
L. Bergstrom and H. Snellman, {Phys. Rev. {\bf D 37} (1988) 3737.}
 
\bibitem{Su}
M. Su and D. P. Finkbeiner, {arXiv:1206.1616 [astro-ph.HE].} 

\bibitem{Bergstrom2}
L. Bergstrom and P. Ullio, {Nucl. Phys. Rev. {\bf B 504} (1997) 27};
P. Ullio and L. Bergstrom, {Phys. Rev. {\bf D 57} (1998) 1962.}

\bibitem{Jackson}
C. B. Jackson, G. Servant, G. Shaughnessy, T. M. P. Tait and M. Taoso, 
{JCAP {\bf 1004} (2010) 004, arXiv:0912.0004 [hep-ph].} 

\bibitem{Landau}
L. D. Landau, {Dokl. Akad. Nawk., USSR {\bf 60} (1948) 207};
C. N. Yang, {Phys. Rev. {\bf 77} (1950) 242.}

\bibitem{Oda}
I. Oda, {Mod. Phys. Lett. {\bf A 27} (2012) 1250116, arXiv:1204.1547 [hep-ph].} 



\end{thebibliography}
\end{document}